\documentstyle[preprint,tighten,aps,epsf]{revtex}

\def\lapprox{\mathrel{\mathop  {\hbox{\lower0.5ex\hbox{$\sim$}
\kern-0.8em\lower-0.7ex\hbox{$<$}}}}}  
\def\gapprox{\mathrel{\mathop  {\hbox{\lower0.5ex\hbox{$\sim$}
\kern-0.8em\lower-0.7ex\hbox{$>$}}}}}

\begin{document}

\draft

\preprint{\vbox{\noindent{}\hfill INFNFE-04-02}}

\title{The Sun and the Newton Constant}

\author{B.~Ricci$^{(1,2)}$ 
and F. L. Villante $^{(1,2)}$ }

\address{
$^{(1)}$Dipartimento di Fisica dell'Universit\`a di Ferrara,I-44100
Ferrara,\\
$^{(2)}$Istituto Nazionale di Fisica Nucleare, Sezione di Ferrara, 
I-44100 Ferrara,}

\maketitle

\begin{abstract}

Several properties of the 
solar interior are determined with a very high  accuracy, which 
in some cases is comparable to that achieved in the determination 
of the Newton constant  $G_N$.
We find that the present uncertainty 
$\Delta G_N/G_N=\pm 1.5\cdot 10^{-3}$ has significant effects on the profile
of density and pressure, however
it has negligible influence on the solar
properties which can be measured by means of helioseismology and
$^8{\rm B}$ neutrinos. Our result do not support recent claims that 
observational solar data can be used to determine the value of
$G_N$ with an accuracy of few part in $10^{-4}$. Present
data  cannot constrain $G_N$ to much better than $10^{-2}$.

\end{abstract}

\section{Introduction}
\label{intro}

In the last ten years, our observational knowledge of the solar 
interior has progressed enormously. By means of helioseismic 
data it has become possible to derive the sound speed 
in the solar interior with   accuracy of  about one part 
per thousand  \cite{eliosnoi,Bahcallelios}.
 By the same method, it has been  possible to deduce 
important properties of the convective envelope. The photospheric  
Helium fraction $Y_{ph}$ and depth of the convective
envelope $R_{b}$  
have been determined with accuracy of about one per cent and one
 per thousand respectively, following the pioneering papers by \cite{Dzie}
and \cite{CD}.
 The measurement of the neutrino flux 
from Boron decay, obtained by combining SNO and Super-Kamiokande 
data \cite{sno,sk,lisi,palazzo},
 has provided a determination of the temperature $T_c$ near 
the solar center with accuracy of about one per cent \cite{what}.

It is impressive that all the predictions of the Standard Solar
Model (SSM) have been confirmed by these accurate tests, 
see e.g. \cite{eliosnoi,Bahcallelios}
and the first rows of table \ref{Taberror} 
for a summary of the available information.

The SSM, like any stellar evolution calculation, depends on several 
parameters. In this respect the Newton constant $G_N$ plays an important 
role, since stellar evolution results from the equilibrium between 
gravitation and other interactions. The sensitivity of stellar 
evolution to modifications of $G_N$ was first stressed by 
Teller \cite{Teller1948}. By means of a homology argument he 
demonstrated that the stellar luminosity is $L_\odot  \propto 
G_{N}^7 M_\odot^5$, so that even a ten per cent variation 
of $G_N$ from its standard value would imply  that life on 
Earth cannot be sustained!

Among the ``fundamental parameters" of nature \cite{Okun,Codata98}, 
 $G_N$ is the most difficult to measure. In fact, it is determined 
with relatively poor accuracy and the situation of  the field 
can be summarized by observing that the CODATA-98 \cite{Codata98}
value, 
$G_N= 6.673(10) 10^{-11}$ m$^3$s$^{-2}$Kg$^{-1}$ includes  a relative accuracy 
$\Delta G_N/G_N=\pm 1.5\cdot 10^{-3}$ 
which is a factor ten larger than that estimated 
in 1986 \cite{Codata86}. Recent experiments \cite{Luo,GM,Quinn} have claimed
 accuracy of about $10^{-5}$ however the disagreement between 
individual results is at the level of $10^{-3}$.

Since several solar properties are now observationally determined
 with an accuracy of order $10^{-3}$,  
we shall address the following questions:

{\em 1) What is the uncertainty on SSM predictions induced by the
uncertainty of $G_N$}?

{\it 2) Can one exploit the available accurate  observations of the 
solar interior, i.e. helioseismic data and Boron neutrino flux,
  for obtaining a determination of $G_N$ with accuracy comparable 
or better than that of laboratory measurements?}

This last question, which is particularly interesting, was recently raised
by Lopes and Silk \cite{Silk}.

We shall present solar models calculated for different value of $G_N$,
deriving the predictions to be compared with  helioseimic 
and neutrino data.  We shall find that  solar sound speeds 
and temperature are actually very weakly dependent on $G_N$ 
and we  shall provide an explanation of this apparently puzzling result.

\section{Solar models}
\label{secsolar}

We have built several solar models corresponding to different
 values of $G_N$, by using an up-to-date version of the evolutionary 
code FRANEC, which includes element diffusion, recent opacity 
tables and modern nuclear reaction rates \cite{Franec}.

For a given value of $G_N$ the three input parameters of the 
code - the mixing length $\alpha$, the initial Helium and metal
 abundances $Y_{in}$ and $Z_{in}$ are varied until one reproduces 
the observed values of the solar radius 
[$R_\odot= (6.9598 (1\pm 0.01\% )10^{8}$ m ],
 luminosity [$L_\odot = 3.844(1 \pm 0.4\%)10^{26}$ W]
 and photospheric composition
 [$Z/X = 0.0245(1 \pm 6\%)$] 
at the solar age [$t_\odot= 4.57 (1\pm 0.4\%)$ Gy].

We remark that the  calculated properties of the solar interior 
are sensitive to the values of these observables, particularly 
to $L_\odot$. For this reason, in order to disentangle the effect of 
tiny changes of $G_N$ we require that $L_\odot$ is fixed 
to the level of $2\cdot 10^{-5}$, i.e. much better than the observational 
accuracy. This precaution is necessary, otherwise the calculated 
models will reflect the changes of $L_\odot$ mixed to the 
changes of $G_N$.

We also remark that astronomical observations fix the product  
$ G_N M_\odot$ quite accurately \cite{stix}:
\begin{equation}
\label{eqgm}
G_N M_\odot = (132\, 712\, 438\,\pm 5 ) 10^{12}\quad m^3s^{-2} \quad.
\end{equation}

Laboratory measurements of $G_N$ are thus measurements of $M_\odot$. 
If $G_N$ is changed
$M_\odot$ has to be varied so that eq.(\ref{eqgm}) is satisfied. 
This is an important point for the present discusssion.

\begin{figure}[tbh]
\epsfysize20truecm
\vskip1cm
\epsfbox{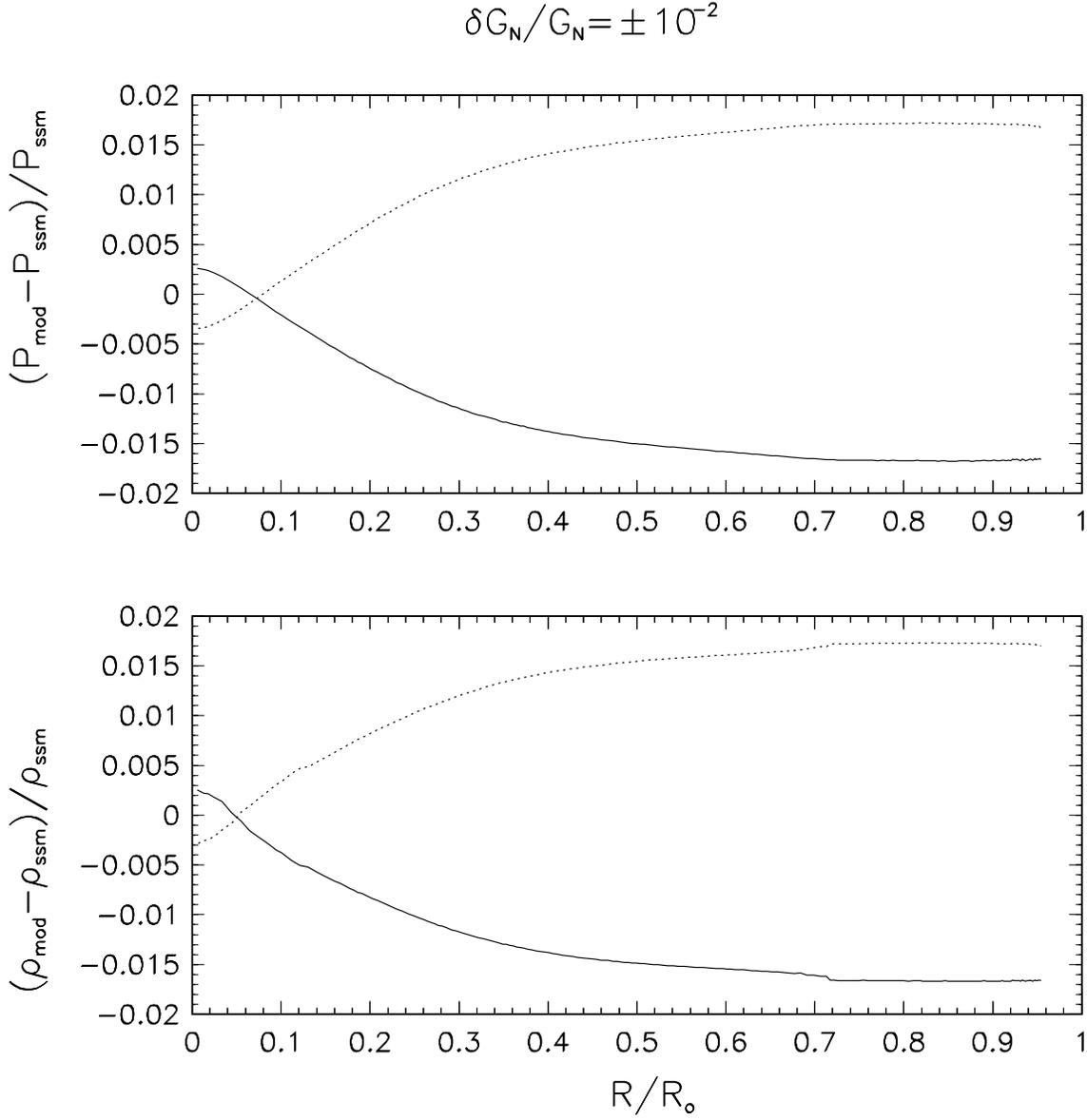}
\caption[c]{Relative change (model-SSM)/SSM  of pressure (upper panel)
and density (lower panel)   as a function 
of the radial coordinate for a change of $G_N$ by +1\% (full line) and 
-1\% (dashed line).}
\label{FigPrho}
\end{figure}

\begin{figure}[tbh]
\epsfysize20truecm
\vskip1cm
\epsfbox{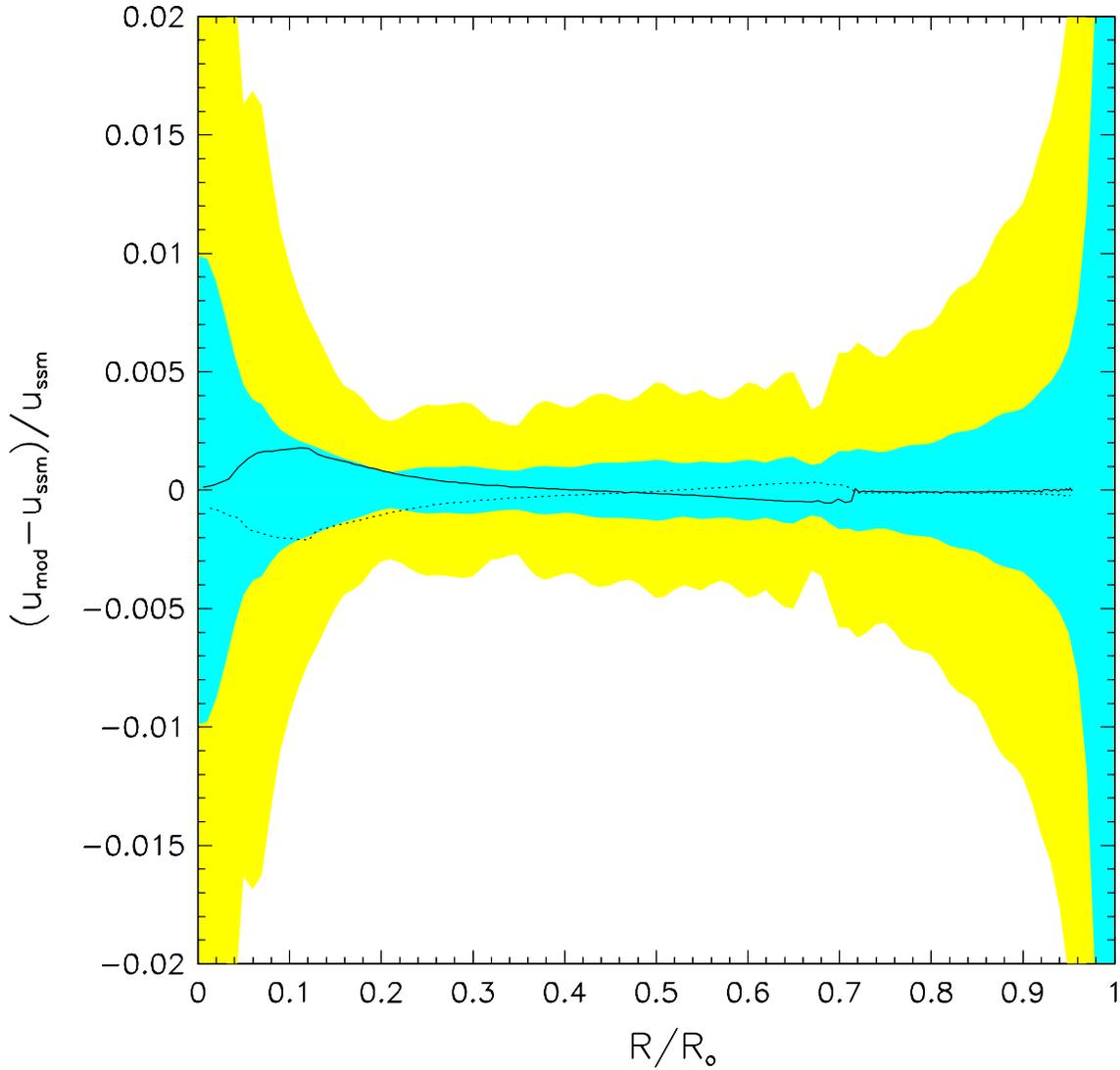}
\caption[a]{Relative change (model-SSM)/SSM  of $u=P/\rho$ as a 
function of the radial coordinate, for a change of $G_N$ by +1\% 
(full line) and -1\% (dotted line). The 
 $1\sigma$ ($3\sigma$) helioseismic uncertainty correspond to 
the dark (light) area.}
\label{Figu}
\end{figure}

\begin{figure}[tbh]
\epsfysize20truecm
\vskip1cm
\epsfbox{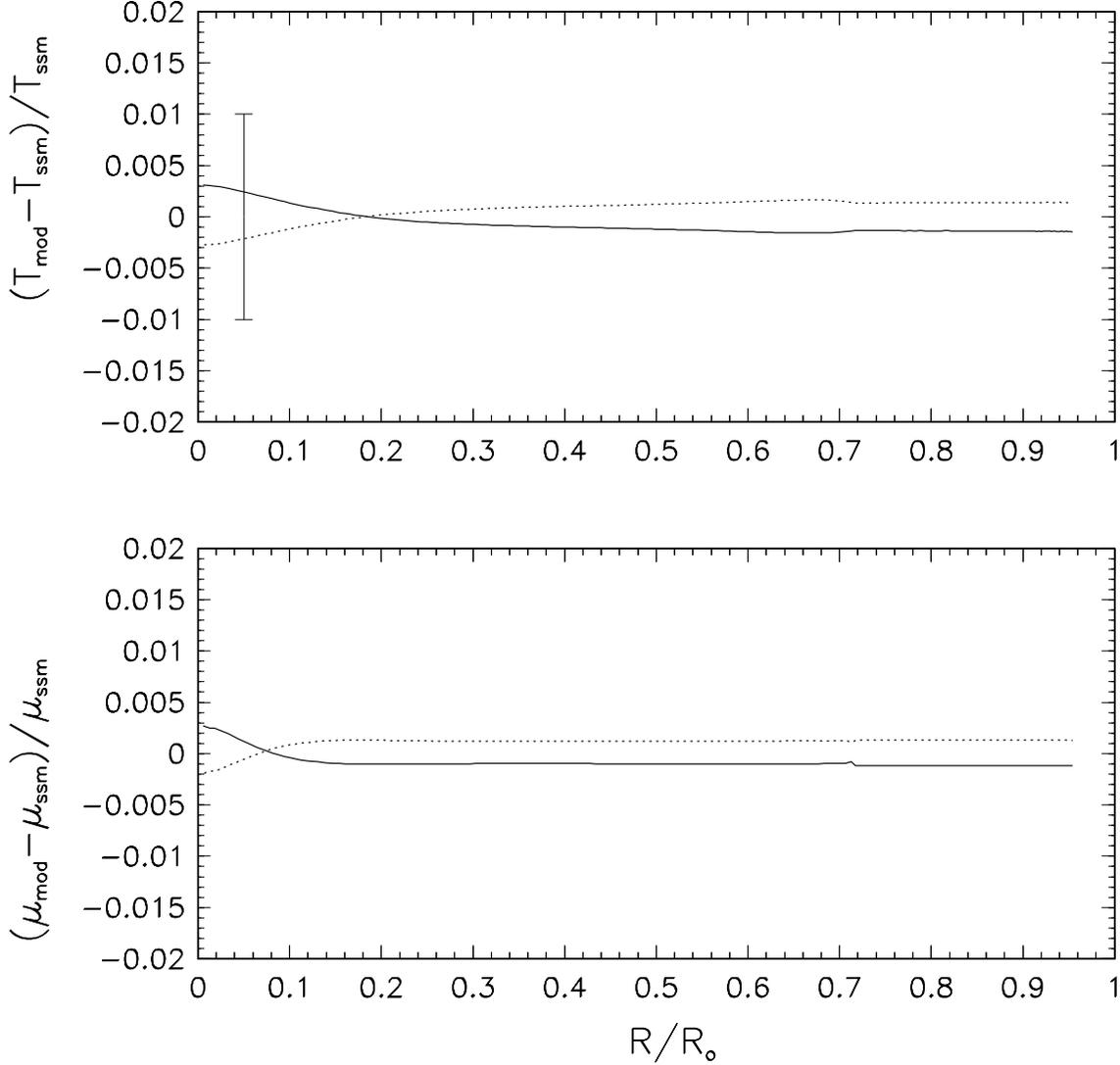}
\caption[b]{Relative change (model-SSM)/SSM  of temperature (upper panel)
and mean molecular weight (lower panel)  as a 
function of the radial coordinate, for a change of $G_N$ by +1\% 
(full line) and -1\% (dashed line). The vertical bar in the upper panel
corresponds to the $1\sigma$ uncertainty on $T_c$ from neutrino measurement.}
\label{FigTmu}
\end{figure}

In Figs. \ref{FigPrho},\ref{Figu},\ref{FigTmu}  
 and in table \ref{Taberror} we present the effect of varying
 $G_N$ by $\pm 1\%$ with respect to the  ``standard" CODATA-98 value. 
The following points are to be remarked:

i) the changes of pressure and density are of the same order of 
$\delta G_N/G_N$, as expected, see Fig. \ref{FigPrho}.

ii) On the other hand 
 the induced variation of the (squared isothermal) sound speed  
$u= P/\rho$ is much smaller. For $\delta G_N/G_N=1\%$ one has
at most $\delta u/u=2\cdot 10^{-3}$, a value comparable to the 
present observational accuracy, see Fig. \ref{Figu}.

iii) Also the variation of temperature  is much suppressed, see
Fig. \ref{FigTmu}: the change of the central temperature is about 0.3\%,
and
that  of the Boron flux is about  5\%, see Table \ref{Taberror}, 
in both cases a factor three below 
the present observational uncertainty.

iv) The depth of the convective zone is altered just by a 
factor $4\cdot 10^{-4}$ and the photospheric helium abundance by less
 than 1\%, well below the present observational uncertainties, see 
again Table \ref{Taberror}.

More generally, we can express the sensitivity of the 
observable $O_{i}$ to the change of $G_N$  by using scaling parameters 
\begin{equation}
\label{eqbeta}
\beta_i = \frac{d\log O_{i}}{ d\log G_N }\quad .
\end{equation}
The calculated  $\beta_i$ values are also presented in Table \ref{Taberror}.
Changes in $G_N$ at the level of present uncertainty ($1.5\cdot 10^{-3}$)
induce changes of helioseismic observables and neutrino fluxes which 
are negligible in comparison with the respective 
observational uncertainty, see last row of Table \ref{Taberror}.

Our results do not support the claim of ref. \cite{Silk} that observational
 solar data  can be used to determine the value of $G_N$  with 
accuracy of few parts in $10^{-4}$.  Variations of $G_N$ at this level 
induce changes which are much  too small in comparison with 
observational accuracy.

{\it In summary, present data are sensitive to changes
of $G_N$ provided that these are of the order of $10^{-2}$.}

\section{Interpretation}
\label{secinterp}

A puzzling situation has emerged. $G_N$ is clearly an  important 
parameter, and its variation induces a comparable change on physically
 relevant quantities such as $P$ and $\rho$.
On the other hand the change of $u$ and $T$ are definitely suppressed.

The reason for the cancellation is in the constancy of $G_N M_\odot$, 
as can be easily demonstrated.

Actually the typical scales  of solar pressure and density are given by:
\begin{eqnarray}
P &=& G_N M_\odot^2/ R_\odot^2\\
\rho &=&  M_\odot/ R_\odot^3
\end{eqnarray}

For a change of $G_N$ which keeps constant the product of 
$G_N M_\odot$ these vary as :
\begin{equation}
\delta P/P  = \delta \rho/ \rho = -\delta G_N / G_N .
\end{equation}
In fact, this is the pattern shown in Fig. \ref{FigPrho}.

In this approximation, when considering  $u=P/\rho$ the effects of 
changing $G_N$ cancel, in other words $u$ is unaffected by changes 
of $G_N$, which explains the much weaker sensitivity emerging 
from the numerical calculations presented in the previous sections.

From the perfect gas law, which describes to a good approximation
 most of the solar core, one has $P/\rho =kT/\mu$  , so that also $T/\mu$ 
is weakly sensitive to changes of $G_N$. The present value of
 the mean molecular weight $\mu$ is mainly determined from the initial 
conditions ($Y_{in}$) and the solar history and not from $G_N$. It follows 
that also $T$ is very weakly sensitive to $G_N$.

{\it We remark that the constancy of  $G_N M_\odot$ is essential for the 
cancellation.} If one computes stellar structures with fixed 
$M_\odot$ and different $G_N$ one finds fractional changes of $u$ and 
$T$ that are proportional to $\delta G_N/G_N$, however these cannot 
be interpreted as solar models.

\section{Future prospects}

One expects that, in the future, the sound speed $u$ will be 
measured  with a higher accuracy, possibly to the level of $10^{-4}$.
Even in this case, however, it will be difficult to get significant 
improvements on $G_N$, unless one achieves 
the same level of accuracy in several other physical 
inputs which affect $u$. 
For instance, let us consider the effect of variations 
of the nuclear reaction cross sections and 
of the opacity, $\kappa$, 
 We remind that these
quantities are affected by uncertainties of few  percent (at least), see
\cite{bp92,nacre,adelberger}.

As we have already remarked, the product $G_N M_\odot$ is
quite accurately fixed, so that if $G_N$ increases, the solar mass 
must decrease according to $\delta M_{\odot}/M_{\odot} = - \delta G_N /G_N$.
A smaller $M_\odot$, being the radius of the sun 
fixed by observational data, implies 
a reduction of number densities $n_i$ of the various  particles
present in the sun.

This implies, a reduction of the nuclear reaction rates
(which are proportional to $n_i n_j$) and, at the same time, an 
increase of the photon mean free path (which, in the
core of the sun, is inversely proportional to 
electron number density). This suggests that suitable 
changes of nuclear energy production rate $\epsilon$
and of opacity can mimic changes of $G_N$. 

\begin{figure}[tbh]
\epsfysize20truecm
\vskip1cm
\epsfbox{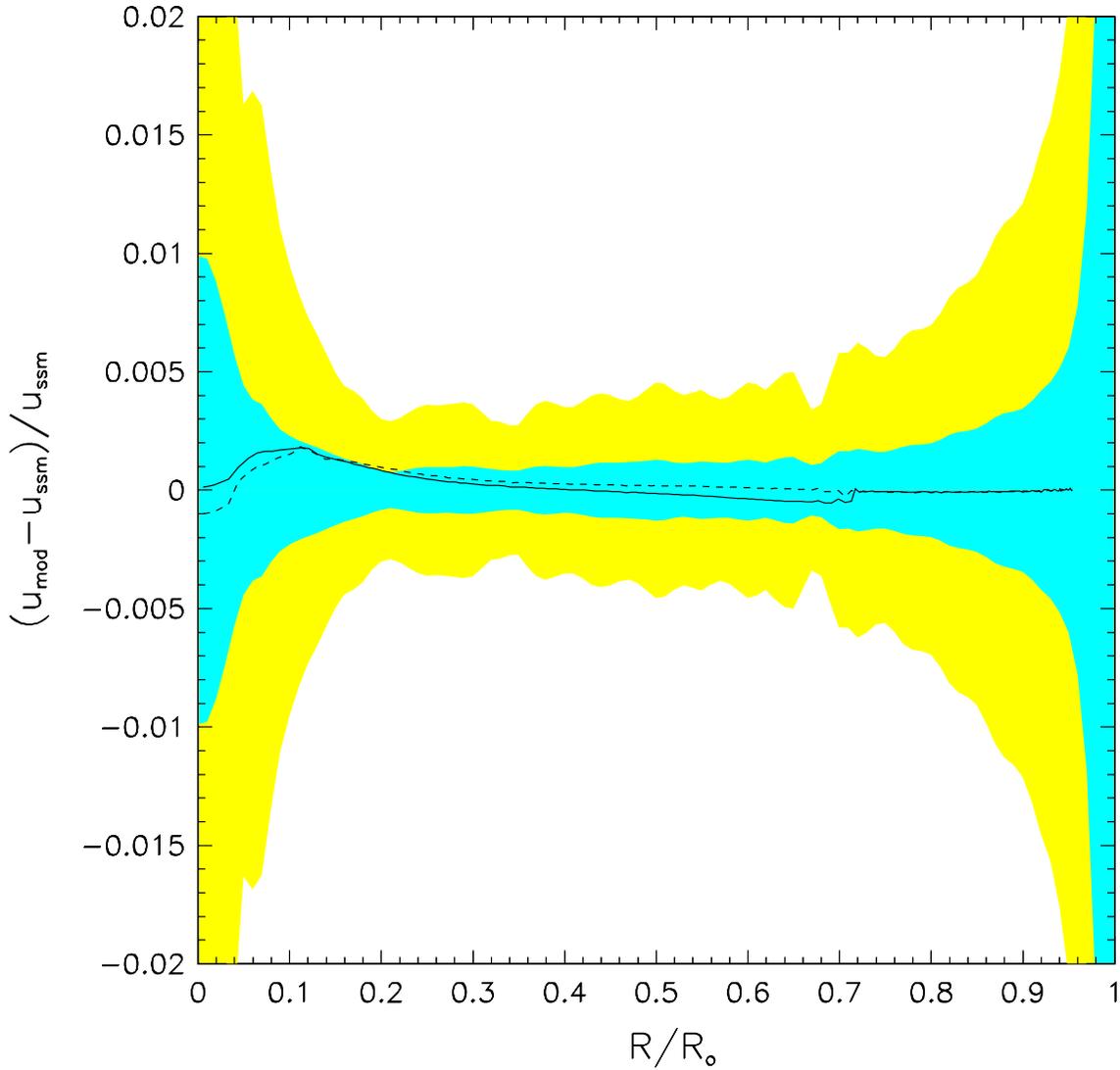}
\caption[a]{Relative change (model-SSM)/SSM  of $u=P/\rho$ as a 
function of the radial coordinate, for a change of $G_N$ by +1\% 
(full line) and for a solar model with energy production rate decreased by
2\% and opacity decreased by 1\% (dashed line). The 
$1\sigma$ ($3\sigma$) helioseismic uncertainty correspond to 
the dark (light) area.}
\label{Figspp}
\end{figure}

Actually, one expects that a combined variation 
$\delta \epsilon/\epsilon = - 2 \delta G_N /G_N$ and
$\delta \kappa / \kappa = - \delta G_N /G_N$ should 
affect the sound speed in the same way as a variation of $G_N$. 
This is supported by the numerical results shown in 
Fig. \ref{Figspp}. We have built a solar model with energy
production rate 
decreased by 2\% and with opacity decreased by 1\%,
and we have obtained a sound speed profile which is similar 
(at the level 0.1\% or less)
to that obtained by increasing $G_N$ by 1\%.

Similar considerations hold for a possible determination
of $G_N$ by means of $\Phi_{\rm B}$ measurements.
By using the $\beta$ coefficient shown in 
Tab.~\ref{Taberror}, one sees that a variation
$\delta G_N/G_N = 10^{-3}$ corresponds to change of
the $^{8} {\rm B}$ neutrino flux 
$\delta\Phi_{{\rm B}} / \Phi_{{\rm B}} = 5\cdot 10^{-3}$
A ``global'' accuracy (i.e. both theorethical and experimental) 
of the order $5\cdot 10^{-3}$ is thus needed in order to 
distinguish among values of $G_N$ which differ by $10^{-3}$.
We remind that the present experimental and theoretical 
determinations of the $^{8} {\rm B}$ neutrino flux have 
uncertainties of the order of 
15-20\% per cent.ns hold for a possible determination
of $G_N$ by means of $\Phi_{\rm B}$ measurements.
By using the $\beta$ coefficient shown in 
Tab.~\ref{Taberror}, one sees that a variation
$\delta G_N/G_N = 10^{-3}$ corresponds to change of
the $^{8} {\rm B}$ neutrino flux 
$\delta\Phi_{{\rm B}} / \Phi_{{\rm B}} = 5\cdot 10^{-3}$
A ``global'' accuracy (i.e. both theorethical and experimental) 
of the order $5\cdot 10^{-3}$ is thus needed in order to 
distinguish among values of $G_N$ which differ by $10^{-3}$.
We remind that the present experimental and theoretical 
determinations of the $^{8} {\rm B}$ neutrino flux have 
uncertainties of the order of 
15-20\% per cent.

\section{Concluding Remarks}

We have found that:

\begin{itemize}

\item
The present uncertainty on the gravitational constant, 
$\Delta G_N/G_N=\pm 1.5 \cdot 10^{-3}$ has significant effects on
the profile of density and pressure.

\item
On the other hand, it has negligible influence on the solar
properties which can be measured by means of helioseismology and
$^8{\rm B}$ neutrinos: sound speed profile, central temperature, depth and helium 
content of the convective envelope.

\item
Our result do not support recent claims \cite{Silk} that 
observational solar data can be used to determine the value of
$G_N$ with an accuracy of few part in $10^{-4}$. Present 
data cannot constrain $G_N$ to much better than $10^{-2}$.
Furthermore, even if the sound speed measurments will become
much more accurate, it will be difficult to get significant improvements
on $G_N$ due to the approximate degeneracy with variations of other 
physical inputs, e.g. nuclear cross sections  and opacity, 
which are presently known at the per cent level or worse.

\end{itemize}

\acknowledgments
We are grateful to G. Fiorentini for useful comments and suggestions.

\begin{table}
\caption[aa]{SSM predictions, observational errors and 
sensitivity to $G_N$. 
$\Delta$ is the $1\sigma$ relative observational error; 
$\delta_{\pm 1\%}$ is the variation
(model-SSM)/model obtained when changing $G_N$ by $\pm 1\%$; 
$\beta$ is the scaling coefficient of eq. \ref{eqbeta};
$\delta_{+0.15\%}$ is the variation
(model-SSM)/model estimated when changing $G_N$ by $+0.15\%$. }
\begin{tabular}{lllllll}
Observable & $u(0.1 R_\odot)$ & $u(0.6 R_\odot)$ &  $T_c$   &  $\Phi_B$ & $Y_{ph}$ & $R_b$\\
units      &  (Mm/s)$^2$      & (Mm/s)$^2$       & $10^7$ K &  $10^6$cm$^{-2}$s$^{-1}$ &  &  $R_\odot$\\
\hline
SSM \cite{BP2000} &0.1525 & 0.09222 & 1.568 & 5.15 &0.2437 &0.7140 \\
$\Delta$ (\%)    &  0.2$^*$& 0.12$^*$ & 1$^{**}$ & 18$^{**}$ & 1.4$^*$ &0.2$^*$ \\
\hline
$\delta_{+1\%} $ (\%) & +0.18  & -0.036& +0.31&+5.7  & -0.67& +0.04\\
$\delta_{-1\%} $ (\%) & -0.20  & +0.021& -0.28&-4.4  & +0.77& -0.04\\
$\beta$               & +0.19  & -0.03& +0.29 &+5.1   &-0.72 & +0.044\\
$\delta_{+0.15\%} $ (\%) & +0.028  & -0.0045  &+0.044&+0.76&-0.11&+0.0065\\
\end{tabular}
{\footnotesize $^*$from helioseismic data \cite{eliosnoi}\\     
               $^{**}$ from $^8$B neutrino data \cite{sno,what}}

\label{Taberror}
\end{table}

\end{document}